\documentstyle[preprint,tighten,aps]{revtex}

\begin{document}
  \draft
  \title{Two phase transitions in the fully frustrated $XY$ model}
  \author{Peter Olsson}
  \address{Department of Theoretical Physics\\
    Ume{\aa} University\\
    901 87 Ume\aa, Sweden\\
    Peter.Olsson@Physics.umu.se}
  \date{\today}
  \maketitle

\begin{abstract}
  The fully frustrated $XY$ model on a square lattice is studied by
  means of Monte Carlo simulations. A Kosterlitz-Thouless transition
  is found at $T_{\rm KT} \approx 0.446$, followed by an ordinary Ising
  transition at a slightly higher temperature, $T_c \approx 0.452$.
  The non-Ising exponents reported by others, are explained as a
  failure of finite size scaling due to the screening length
  associated with the nearby Kosterlitz-Thouless transition.
\end{abstract}


The critical behavior of two-dimensional fully frustrated $XY$
(FF$XY$) models has been the subject of much interest during the last
decade.  This is seen in the large number of papers in the
literature, of which a number are very recent.  In spite of this, the
controversy of the nature of the phase transition(s) in these models
is by no means settled.

The models under discussion include the anti-ferromagnetic $XY$ model
on a triangular lattice \cite{Miyashita_Shiba,LJN_Landau}, the
square-lattice version of the $XY$ model with one anti-ferromagnetic
coupling per plaquette \cite{Villain:77,Teitel_Jayaprakash:83a}, and the
corresponding Coulomb gas with half-integer charges
\cite{Thijssen_Knops:88,J-R.Lee:94}.  Also discussed in this context are
the coupled $XY$-Ising system \cite{G_Kosterlitz_LN} and the 19-vertex
version of the fully frustrated $XY$ model \cite{Knops_NKB}, which are
believed to be in the same universality class.  In the present Letter
we focus on the square-lattice version of the FF$XY$\  model, but since
the results are expected to have a more general validity, we
will repeatedly refer to studies of the other above-mentioned models.

Beside the theoretical questions regarding the universality class, the
study of the fully frustrated models is largely motivated by their
relevance for Josephson junction arrays in a magnetic field.  Due to
this relation, the Hamiltonian of the FF$XY$\  model on a square lattice
is customarily written with the vector potential $A_{ij}$,
\begin{displaymath}
  H = -J\sum_{\left<ij\right>} \cos(\theta_i - \theta_j + A_{ij}).
\end{displaymath}
Here $i$ and $j$ enumerate the lattice sites, $\theta_i$ is an angle
associated with site $i$, and the sum is over nearest neighbors.  The
frustration is determined by the $A_{ij}$. Full frustration
corresponds to one half flux quantum per plaquette, which means that
$f \equiv \frac{1}{2\pi}\sum A_{ij} = 1/2$, where the sum is taken
around a plaquette, cf.\ Eq.\ (\ref{def.m}) below.

The ground state for this model on a square lattice \cite{Villain:77},
has plaquettes with clockwise and counter-clockwise rotation in a
checkerboard pattern. The angular difference between nearest neighbors
is $\phi_{ij} \equiv \theta_i - \theta_j + A_{ij} = \pm \pi/4$. This
checkerboard pattern gives rise to the discrete $Z_2$ symmetry of the
anti-ferromagnetic Ising model, beside the rotational $XY$ symmetry.
At low temperatures this model therefore has both the topological
long-range order of the $XY$ model, and the ordinary long-range order
of the anti-ferromagnetic Ising model. As the temperature is
increased, both the $XY$-like and the Ising-like orders are expected to
vanish.

In the first Monte Carlo (MC) study of this model, Teitel and
Jayaprakash \cite{Teitel_Jayaprakash:83a} found a steep drop in the
helicity modulus, signaling the loss of $XY$ order, accompanied by an
increase in the specific heat with lattice size, consistent with an
Ising transition. Being unable to determine the precise critical
behavior of the FF$XY$\  model, the authors put forward two possible
scenarios:
\begin{itemize}
\item[(i)] As the Ising temperature, $T_c$, is approached from below,
  the Ising excitations produce a steep drop in the helicity modulus.
  As this quantity approaches the universal value, the KT excitations
  become important producing a universal jump.  This occurs before the
  loss of Ising order, $T_{\rm KT} < T_c$.
\item[(ii)] As $T_c$ is approached from below, the Ising excitations
  give rise to a jump larger than the universal value
  \cite{Minnhagen:85c}.  The loss of Ising and $XY$ order take place
  at the same temperature, $T_{\rm KT} = T_c$.
\end{itemize}
Since then, several investigations have been made with the aim to
decide between these two possibilities. Whereas some of the
earliest MC studies not were decisive
\cite{Miyashita_Shiba,LJN_Landau}, a large number of recent papers
\cite{J-R.Lee:94,G_Kosterlitz_LN,Knops_NKB,Thijssen_Knops:90,%
L_Kosterlitz_G,Granato_Nightingale,Lee_Lee,Ramirez-Santiago_Jose:94}
on FF$XY$\  models, have yielded exponents for the $Z_2$ transition which
differ from the pure Ising ones, suggesting a new universality
class, the second possibility above.  The evidence is, however, not
conclusive, since the finite size scalings in these papers not are
quite satisfactory.

In this Letter we present some MC analyses that shed new
light on the behavior of the FF$XY$\  models. We first give evidence for
an ordinary KT transition.  We then demonstrate that the presence of
the screening length associated with this transition in the region
immediately above $T_{\rm KT}$, precludes the use of ordinary finite size
scaling, and argue that this is the reason for the reported non-Ising
exponents. We then determine the $Z_2$ correlation length, present
evidence for an Ising temperature $T_c > T_{\rm KT}$, and demonstrate that
our data are, indeed, consistent with the pure Ising exponent, $\nu=1$.
Our MC data is obtained on a bunch of workstations, by means of the
ordinary Metropolis algorithm.

The precise determination of the temperature for a Kosterlitz-Thouless
transition is a difficult task.  This is due both to the absence of
spectacular peaks, and a logarithmic correction that gives problems
with ordinary finite size scaling.  A way to cope with
these difficulties, by extracting the finite-size dependence from the
Kosterlitz' renormalization group equations, was suggested some years
ago \cite{Weber_Minnhagen:88}.  The result may be expressed as a
finite size scaling relation for the helicity
modulus \cite{Ohta_Jasnow} valid right at the transition temperature
$T_{\rm KT}$,
\begin{equation}
  \label{weber-scaling}
  \frac{\Upsilon_L\;\pi}{2T_{\rm KT}} = 1 + \frac{1}{2(\ln L + l_0)}.
\end{equation}
Here $L$ is the system size, $\Upsilon_L$ is the helicity modulus for
that size, and $l_0$ is a parameter to be determined.
The successful application of this relation does, however, require
some care.  In the $XY$ model the amplitude of the spin waves change
with both temperature and lattice size, which makes it necessary to
identify the temperature scale of relevance for the vortices, the
Coulomb gas temperature, $T^{\rm CG}$ \cite{Olsson:Vl,Olsson:Kost-fit}.  In
the Coulomb gas this is no problem, but to obtain the equivalent of
the helicity modulus, one has to include another term in the
Hamiltonian containing the polarization squared
\cite{Vallat_Beck,Olsson:self-cons.long}. The data obtained in this way
does, indeed, fit very well to Eq.\ (\ref{weber-scaling})
\cite{Olsson_Wallin}.

Figure \ref{fig-Ucorr} shows the result from this kind of fit for the
FF$XY$\  model. Note that the scaling relation is obeyed only for rather
large lattices, $L\geq 32$. This is not surprising since Eq.\
(\ref{weber-scaling}) is expected to be valid only for low
renormalized vortex density. A KT transition at a low Coulomb gas
temperature \cite{J-R.Lee:94}, as in the model under consideration,
has to be monitored at larger lengths in order to get data from the
region sufficiently close to the critical point. The fit gives
$T^{\rm CG}_{\rm KT}= 0.12847(4)$ corresponding to $T_{\rm KT}/J = 0.4460(1)$.
We note that this is clearly below $T/J \approx 0.454$, which is
a typical value of the $Z_2$ transition temperature in the
literature \cite{L_Kosterlitz_G,Granato_Nightingale,Lee_Lee}.

For the study of the $Z_2$ transition it is customary to define the
staggered magnetization
\begin{equation}
  M = \frac{1}{L^2} \left| \sum_{\bf r} (-1)^{r_x + r_y} m_{\bf r} \right|,
  \label{def.M}
\end{equation}
where the sum is over all the plaquettes of the system, and $m$ is the
vorticity.  We define the vorticity, in terms of the rotation of
the current $\sin\phi_{ij} \equiv \sin( \theta_i - \theta_j + A_{ij})$
around a plaquette \cite{Miyashita_Shiba},
\begin{equation}
  m = \frac{1}{\sqrt{8}}(\sin\phi_{12} + \sin\phi_{23} +  \sin\phi_{34}
  + \sin\phi_{41}).
  \label{def.m}
\end{equation}
The normalization factor in Eq.\ (\ref{def.m}) is chosen from the
zero-temperature value of $m$, which follows from the angular
difference $\phi=\pm\pi/4$ in the ground state.

Several recent MC analyses of the $Z_2$ transition in the FF$XY$\  models
make use of the expected finite size dependence of various quantities
at criticality \cite{J-R.Lee:94,G_Kosterlitz_LN,Knops_NKB,%
Thijssen_Knops:90,L_Kosterlitz_G,Granato_Nightingale,Lee_Lee}.
Such methods have generally yielded non-Ising exponents, suggesting a
single transition in a new universality class.  One of several
different approaches is to make use of properties of the distribution
function of $M$ at criticality.  This has been done both directly in
the FF$XY$\  model \cite{Lee_Lee} and in the corresponding Coulomb gas
\cite{J-R.Lee:94}.

The tacit assumption behind these scaling analyses is, however, that
the system size is the only relevant length at criticality. The
presence of a nearby transition with a corresponding characteristic
length, $\lambda$, may well invalidate this assumption.  This
therefore calls in question the attempts to determine the critical
exponents of the $Z_2$ transition in the FF$XY$\  models by finite size
scaling at $T_c$.  The condition for a successful application of
finite-size scaling would be $L\gg \lambda$, which may imply
prohibitively large lattices.

The effect of this additional length is clearly seen in the dependence
of $M$ on the boundary conditions (BC).  Beside the ordinary periodic
BC (PBC) we make use of fluctuating BC (FBC) \cite{Olsson:self-cons}
obtained by introducing phase mismatches across the boundaries in the
$x$ and $y$ directions as a pair of additional dynamical variables. It
is with these BC that the $XY$ model corresponds to the CG with
periodic BC
\cite{Thijssen_Knops:88,Olsson:self-cons,Olsson:self-cons.long}.
Results for the magnetization obtained with these two BC, and $L =
16$, 64, are shown in Fig.\ \ref{fig-pfbc}. As the figure shows there
is a size-dependent temperature region where $M$ is sensitive to the
BC.  A comparison with the helicity modulus (dashed lines), gives at
hand that the difference between the two curves vanishes when
$\Upsilon_L \approx 0$.  This condition implies $L \gg \lambda$ since
the vanishing of $\Upsilon$ means that vortex pairs at distance
$\approx L$ are free.

The dependence of $M$ on the BC is presumably a reflection of the
dependence of the vortex interaction on the BC
\cite{Olsson:self-cons.long}, an effect that vanishes if $\Upsilon
=0$. The conclusion from this figure is therefore that the value of
$M$, at fixed $T$, is uniquely determined by the system size, only if
$L \gg \lambda$.

The implications for ordinary finite size scaling is illustrated with
Binder's cumulant \cite{Binder:cumulant},
\begin{displaymath}
  U = 1 - \frac{\left<M^4 \right>}{3\left<M^2 \right>^2}.
\end{displaymath}
For the usual cases where finite size scaling works, this quantity
is size-independent at the critical point, $U_L = U^*$. Both
the critical temperature and the exponent $\nu$, are then obtained from
the crossing of $U_L$ for different $L$.  Figure
\ref{fig-binder} shows this kind of analysis for the FF$XY$\  model with
both PBC and FBC, upper and lower points, respectively. A determination of
$T_c$ from data obtained with PBC, would give $T_c \approx 0.454J$,
in agreement with previous results
\cite{L_Kosterlitz_G,Granato_Nightingale,Lee_Lee}. About the same
critical temperature is obtained from the data obtained with FBC.
In both cases there is, however, no unique crossing
point independent of the lattice sizes.

Beside the lack of an unique crossing point, the problem with this
kind of analysis is that data for the two different BC suggest
different values of $U^*$. It is, however, clear that there can be
only one correct value for $U^*$.  This follows since $U_L$ for
sufficiently large $L$ will be independent of the BC.

Our conclusions so far for the $Z_2$ transition are, firstly, that the
reported non-Ising exponents are artifacts due to the screening length
that invalidates the scaling assumption, and, secondly, that in order
to obtain reliable data for the analysis of the $Z_2$ transition one
has to make use of systems with $L\gg\lambda$, or, equivalently,
$\Upsilon \approx 0$.

For the determination of the $Z_2$ correlation lengths around $T_c$,
we introduce the correlation function
\begin{equation}
  g({\bf r}) =  (-1)^{r_x + r_y} \left<m_0 m_{\bf r} \right>,
\end{equation}
where the prefactor takes care of the anti-ferromagnetic structure.

Below $T_c$ this function decays to a finite value in the large-$r$
limit, $g(r) \rightarrow M^2$.  The decay to this constant is governed
by the correlation length $\xi_-$,
\begin{equation}
  g(r) - M^2 \sim e^{-r/\xi_-}.
  \label{g.low}
\end{equation}
Figure \ref{fig-xi.TKT} shows $g(r)$ at $T_{\rm KT}$. The data is for both
PBC and FBC, at lattices of size $L = 64$, 128.  From the experience
with these BC \cite{Olsson:self-cons} we expect $g(r)$
for an infinite system to lie in between the values for the largest
lattices. The solid line in Fig.\ \ref{fig-xi.TKT} is from Eq.\
(\ref{g.low}) with $M^2 = 0.376$ and $\xi_- = 6.1$, giving strong
evidence that the $Z_2$ transition takes place at a higher
temperature, $T_c > T_{\rm KT}$.

Granted the existence of two separate transitions, one certainly
expects the pure Ising exponents for the $Z_2$ transition. To verify
that expectation by MC simulations, we turn to the
temperature-dependence of the correlation length above $T_c$.  In that
temperature region $M$ vanishes, and the correlation function $g(r)$
behaves as
\begin{equation}
  g(r) \sim e^{-r/\xi}.
  \label{g.xi}
\end{equation}
Again, we make use of data for lattices and temperatures such
that $\Upsilon \approx 0$, which also turns out to be a prerequisite
for a good fit to Eq.\ (\ref{g.xi}). This means that we are only able
to obtain reliable values for $\xi$ at temperatures down to $T/J =
0.472$. As an additional complication, we also have to consider
effects from the temperature dependence of the spin waves.

The effect of the spin waves on the behavior of the FF$XY$\  model seems
to be overlooked so far. In the analysis of the $Z_2$ transition, the
importance of the spin waves stems from the fact that the energy
associated with a domain wall becomes smaller with larger spin wave
amplitude.  There are two possible contributions to the
temperature-dependence: The average value of $|m|$ for a single
plaquette changes with temperature, and the bare vortex interaction
may be affected by the spin waves.

To appreciate the significance of this effect one may well compare
with an Ising model, $H_I = -K\sum_{\left<ij \right>} s_i s_j$, with
non-singular temperature-dependences in both the coupling constant and
the magnitude of the spins.  In the immediate vicinity of $T_c$ these
temperature dependences may be neglected with impunity,
but with data in a larger temperature region, one has to resort to an
effective temperature variable, $T^I = T/(Ks^2)$.  To apply this kind
of reasoning to the FF$XY$\  model we suggest making use of the
expression for the bare vortex interaction from Ref.\
\onlinecite{Olsson:self-cons.long}.  With $m$ defined as in Eq.\
(\ref{def.m}), the energy associated with the vorticity at the origin
and ${\bf r}$ becomes
\begin{displaymath}
  E_{\rm bare}({\bf r}) = \frac{8J^2}{J_0} m_0 G({\bf r}) m_{\bf r},
\end{displaymath}
where $J_0 = J \left<\cos\phi \right>$.  This suggests that the
relevant temperature scale is
\begin{displaymath}
  T^I = \frac{T}{ \left<|m|\right>^2 J/J_0}.
\end{displaymath}

Figure \ref{fig-xi.T} shows $1/\xi$ plotted against $T^I$. This is a
demonstration that the temperature dependence of $\xi$ is, indeed,
consistent with the pure Ising exponent, $\nu = 1$.  The analysis also
suggests a value of the Ising temperature.  The figure gives $T^I_c/J
= 0.4576(13)$, slightly above $T_{\rm KT}^I/J \approx 0.440$.  In ordinary
temperatures this corresponds to $T_c/J = 0.452(1)$, which is
consistent both with $T_{\rm KT}$ as a lower bound, and the crossing points
in Fig.\ \ref{fig-binder} as upper bounds for $T_c$.

In conclusion, we have found ample evidence for two distinct
transitions in the FF$XY$\  model on a square lattice. The KT transition
is analyzed by finite size scaling of the helicity modulus, the
previously obtained non-Ising exponents are explained as a failure of
the scaling assumption, and, with the identification of the relevant
temperature scale, the data is consistent with the Ising exponent,
$\nu = 1$.

The author thanks Professor Petter Minnhagen and Mats Wallin for
critical reading of the manuscript. Financial support from the Swedish
Natural Research Council through Contract No.\ E-EG 10376-302 is
gratefully acknowledged.

\begin{figure}
  \caption{$\Upsilon\pi/2T$ versus lattice size at $T^{\rm CG}_{\rm KT} =
    0.12847$ together with the finite size scaling function, Eq.\
    (\protect\ref{weber-scaling}).  The good fit is strong evidence
    for an ordinary KT transition.}
  \label{fig-Ucorr}
\end{figure}

\begin{figure}
  \caption{Staggered magnetization, $M$, with different BC. Data for
    PBC lie above the corresponding data for FBC.  The dashed lines
    are the helicity modulus for the same sizes.  The data differ at
    temperatures where $\Upsilon \neq 0$.}
  \label{fig-pfbc}
\end{figure}

\begin{figure}
  \caption{Binders cumulant as a function of temperature for the two
    different BC and several lattice sizes.  The difference between
    data from PBC and FBC (solid and dashed lines, respectively) casts
    doubt on the results from this kind of analysis.  }
  \label{fig-binder}
\end{figure}

\begin{figure}
  \caption{Correlation function at $T_{\rm KT}$. Data is for PBC and FBC,
    upper and lower points. The finite values at large
    $r$ shows that this temperature is below $T_c$.}
  \label{fig-xi.TKT}
\end{figure}

\begin{figure}
  \caption{The dependence of the $Z_2$ correlation length on the effective
    temperature. With $\xi^{-1/\nu} \propto T^I - T^I_c$, the data is
    consistent with the pure Ising exponent $\nu = 1$. }
  \label{fig-xi.T}
\end{figure}


\begin{thebibliography}{10}

\bibitem{Miyashita_Shiba}
S. Miyashita and H. Shiba, J.\ Phys.\ Soc.\ Jpn. {\bf 53},  1145  (1984).

\bibitem{LJN_Landau}
D.~H. Lee, J.~D. Joannopoulos, J.~W. Negele, and D.~P. Landau, Phys. Rev. B
  {\bf 33},  450  (1986).

\bibitem{Villain:77}
J. Villain, J. Phys.\ C {\bf 10},  1717  (1977).

\bibitem{Teitel_Jayaprakash:83a}
S. Teitel and C. Jayaprakash, Phys. Rev. B {\bf 27},  598  (1983).

\bibitem{Thijssen_Knops:88}
J.~M. Thijssen and H.~J.~F. Knops, Phys. Rev. B {\bf 37},  7738  (1988).

\bibitem{J-R.Lee:94}
J.-R. Lee, Phys. Rev. B {\bf 49},  3317  (1994).

\bibitem{G_Kosterlitz_LN}
E. Granato, J.~M. Kosterlitz, J. Lee, and M.~P. Nightingale, Phys. Rev. Lett.
  {\bf 66},  1090  (1991).

\bibitem{Knops_NKB}
Y.~M.~M. Knops, B. Nienhuis, H.~J.~F. Knops, and H.~W.~J. Bl\"ote, Phys. Rev. B
  {\bf 50},  1061  (1994).

\bibitem{Minnhagen:85c}
P. Minnhagen, Phys. Rev. B {\bf 32},  7548  (1985).

\bibitem{Thijssen_Knops:90}
J.~M. Thijssen and H.~J.~F. Knops, Phys. Rev. B {\bf 42},  2438  (1990).

\bibitem{L_Kosterlitz_G}
J. Lee, J.~M. Kosterlitz, and E. Granato, Phys. Rev. B {\bf 43},  11531
  (1991).

\bibitem{Granato_Nightingale}
E. Granato and M.~P. Nightingale, Phys. Rev. B {\bf 48},  7438  (1993).

\bibitem{Lee_Lee}
S. Lee and K.-C. Lee, Phys. Rev. B {\bf 49},  15184  (1994).

\bibitem{Ramirez-Santiago_Jose:94}
G. Ramirez-Santiago and J.~V. Jos\'{e}, Phys. Rev. B {\bf 49},  9567  (1994).

\bibitem{Weber_Minnhagen:88}
H. Weber and P. Minnhagen, Phys. Rev. B {\bf 37},  5986  (1988).

\bibitem{Ohta_Jasnow}
T. Ohta and D. Jasnow, Phys. Rev. B {\bf 20},  139  (1979).

\bibitem{Olsson:Vl}
P. Olsson, Phys. Rev. B {\bf 46},  14598  (1992).

\bibitem{Olsson:Kost-fit}
P. Olsson, Phys. Rev. B {\bf 52},  in press  (1995).

\bibitem{Vallat_Beck}
A. Vallat and H. Beck, Phys. Rev. B {\bf 50},  4015  (1994).

\bibitem{Olsson:self-cons.long}
P. Olsson, Phys. Rev. B {\bf 52},  in press  (1995).

\bibitem{Olsson_Wallin}
P. Olsson and M. Wallin (unpublished).

\bibitem{Olsson:self-cons}
P. Olsson, Phys. Rev. Lett. {\bf 73},  3339  (1994).

\bibitem{Binder:cumulant}
K. Binder, Phys. Rev. Lett. {\bf 47},  693  (1981).

\end{thebibliography}
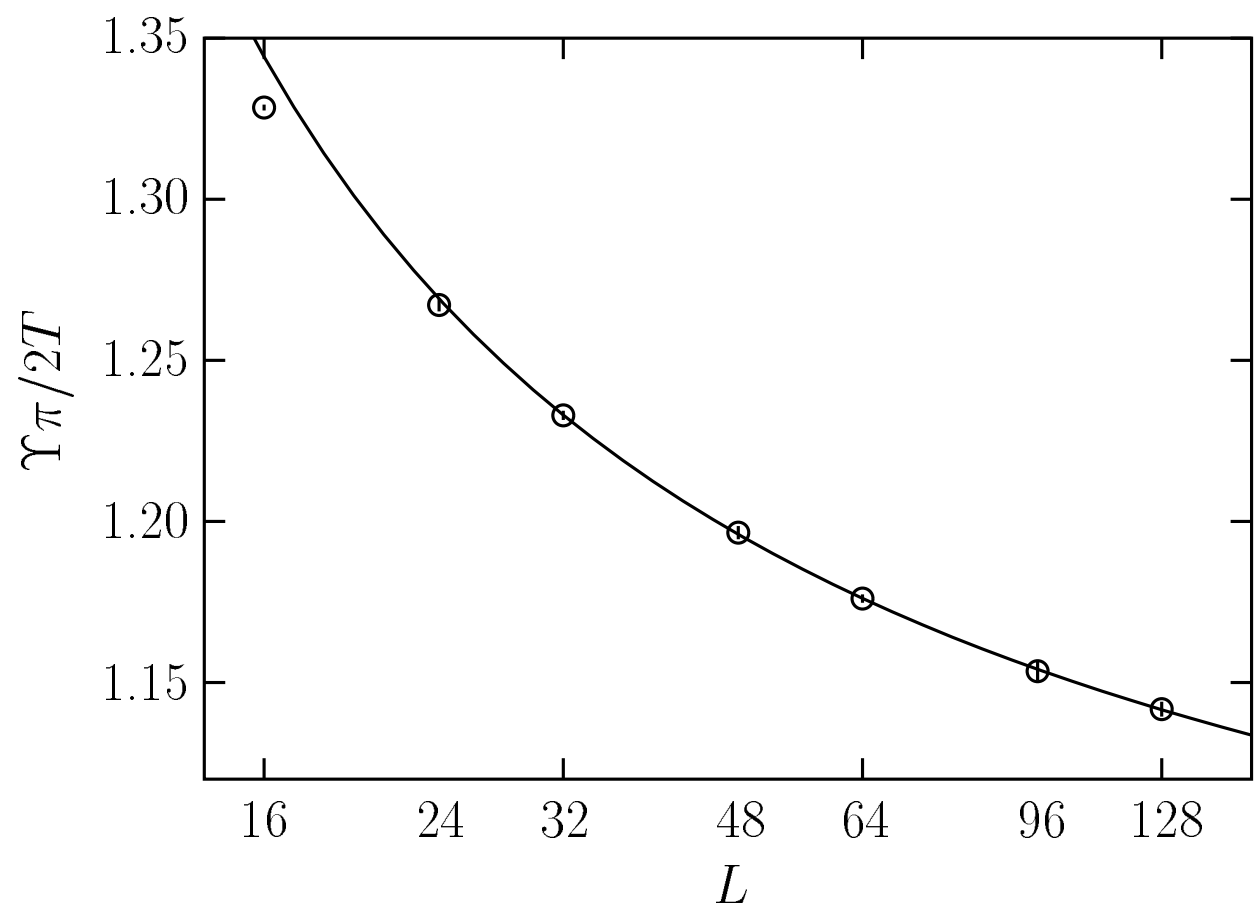
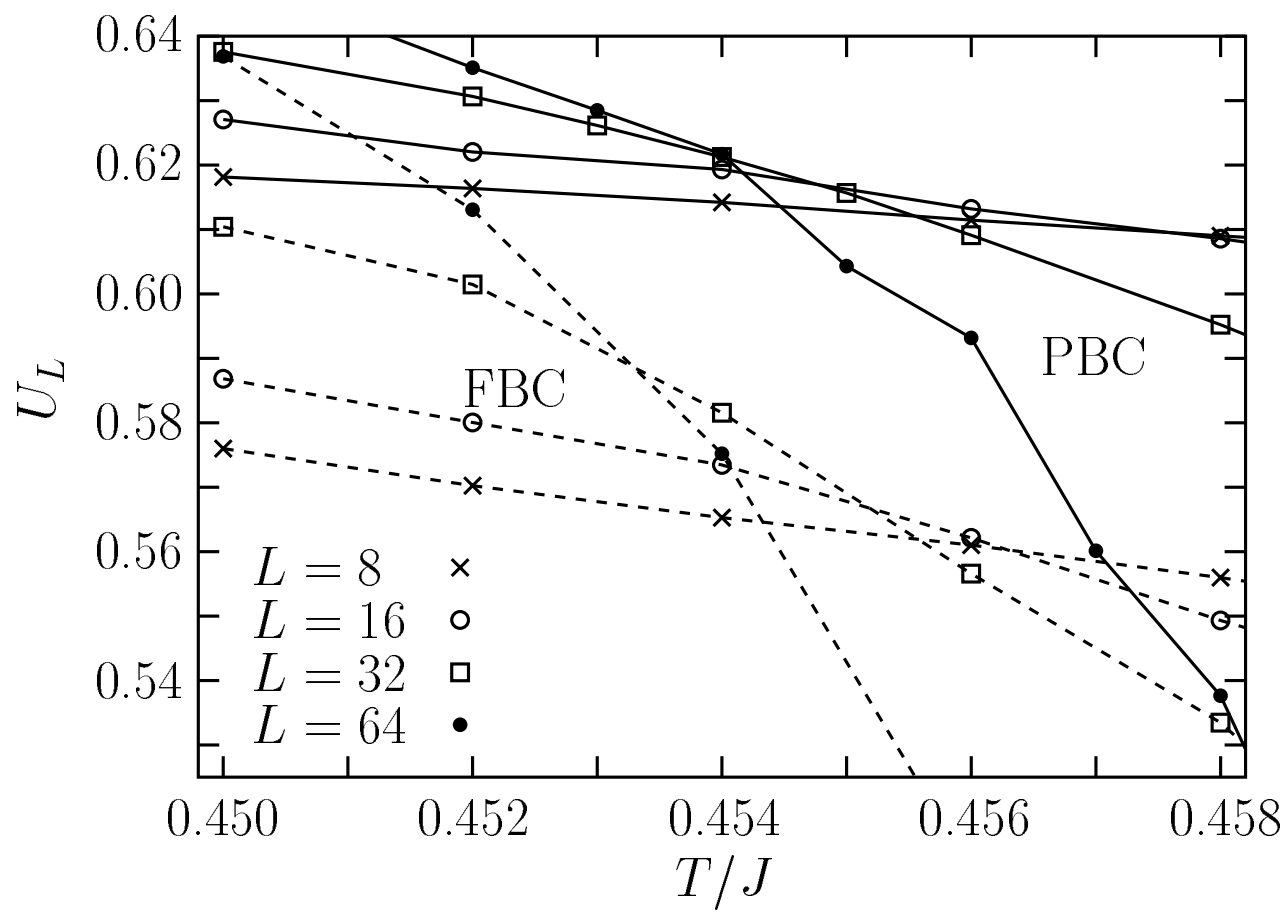
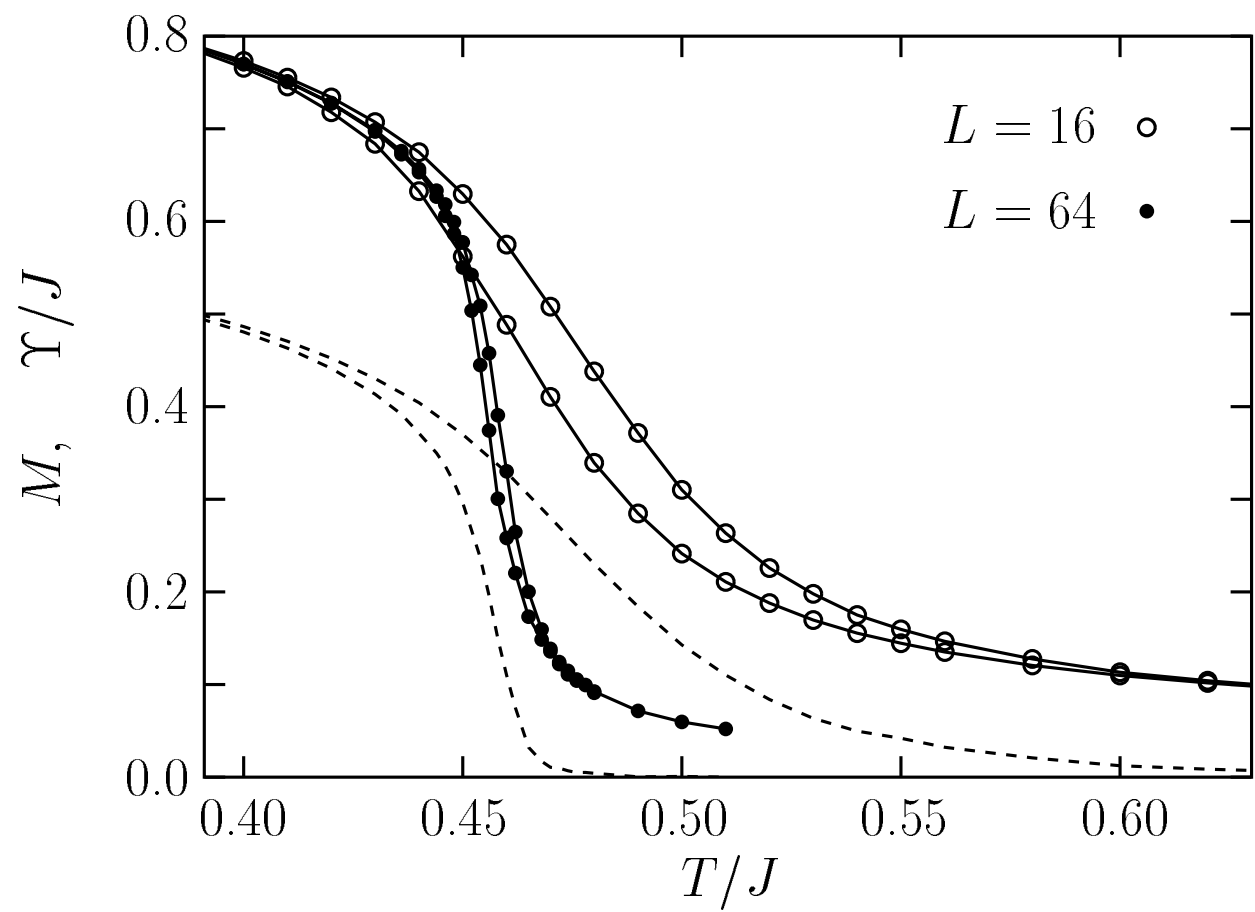
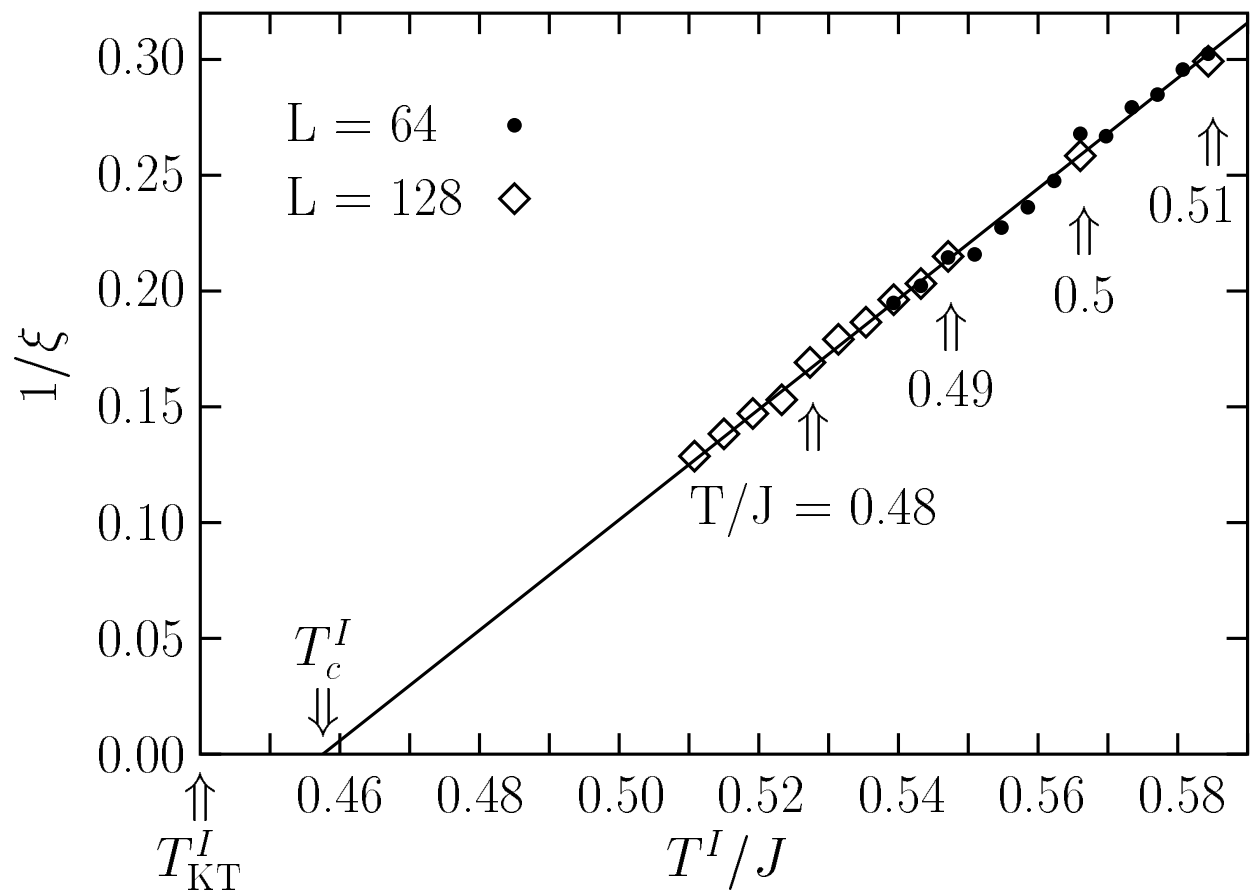
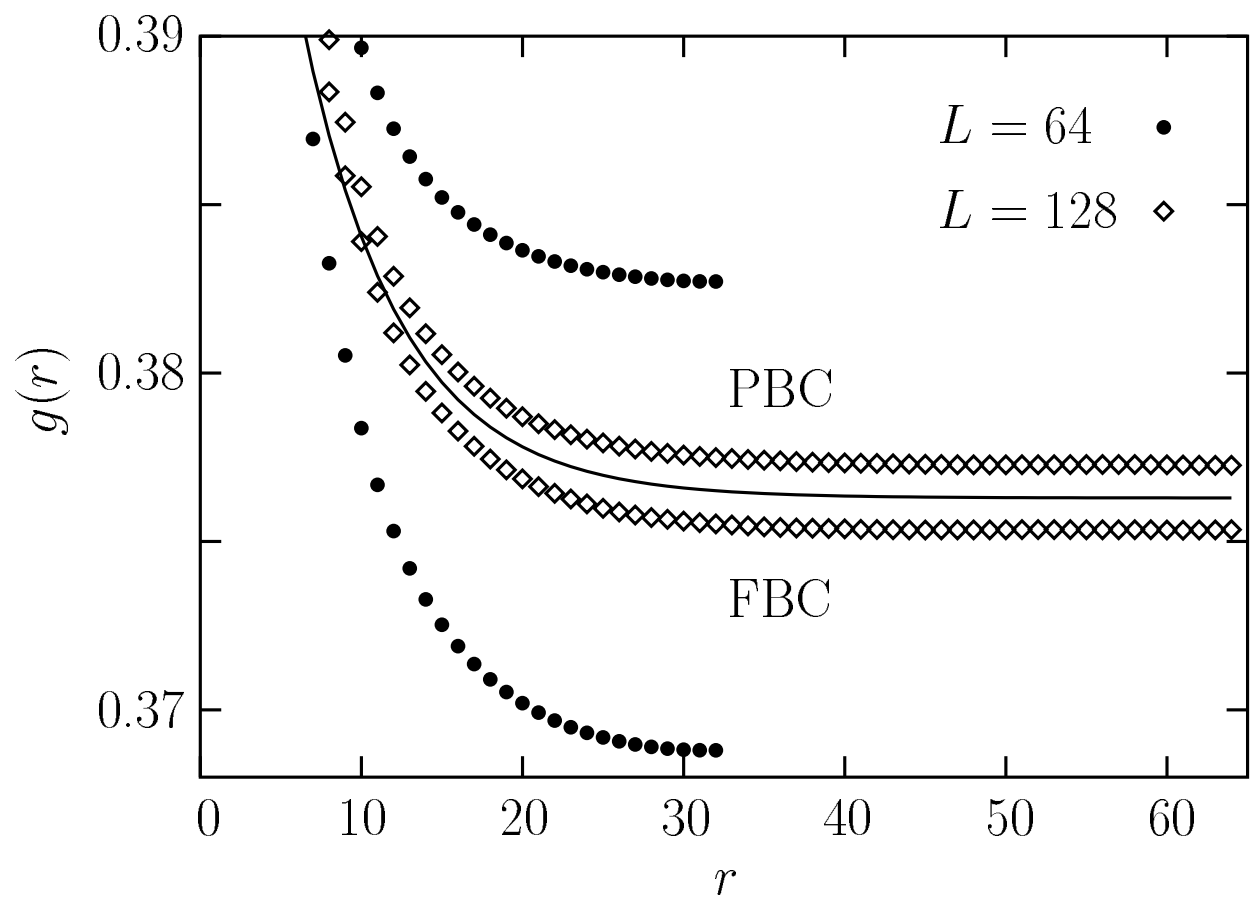
\end{document}